\documentclass[aps,reprint,showpacs,amsmath,amssymb,elsarticle,floatfix,nofootinbib]{revtex4-1}
\usepackage{color}
\usepackage{graphicx}% Include figure files
\usepackage{dcolumn}% Align table columns on decimal point
\usepackage{bm}% bold math
\usepackage{rotating}
\usepackage{float}
\usepackage{xcolor}
\usepackage{natbib}
\usepackage{color}
\usepackage{times}
\usepackage[utf8x]{inputenc}
\usepackage[colorlinks,citecolor=blue,linkcolor=red,anchorcolor=blue,filecolor=blue,urlcolor=blue]{hyperref}
\usepackage{comment}
\begin{document}
%%%
\title{Systematic study of surface properties for Ne, Na, Mg, Al and Si isotopes in a coherent density fluctuation model using the relativistic mean field formalism}

\author{Jeet Amrit Pattnaik$^{1}$}
\email{jeetamritboudh@gmail.com}
\author{R. N. Panda$^{1}$}
\email{rabinarayanpanda@soa.ac.in}
\author{M. Bhuyan$^{2,3}$}
\email{bunuphy@yahoo.com}
\author{S. K. Patra$^{4,5}$}
\email{patra@iopb.res.in}
%%%%%%%%%%%%%%%%%%%%%%%%%%%%%%%%%%%%%%

\affiliation{$^1$Department of Physics, Siksha $'O'$ Anusandhan, Deemed to be University, Bhubaneswar-751030, India}
\affiliation{$^2$Center of theoretical and Computational Physics, Department of Physics, University of Malaya, Kuala Lumpur, 50603, Malaysia}
\affiliation{$^3$Institute of Research and Development, Duy Tan University, Da Nang 550000, Vietnam}
\affiliation{$^4$Institute of Physics, Sachivalya Marg, Bhubaneswar-751005, India}
\affiliation{$^5$Homi Bhabha National Institute, Training School Complex, Anushakti Nagar, Mumbai 400094, India}
\date{\today}
\begin{abstract}
We have systematically studied the surface properties, such as symmetric energy, neutron pressure, and symmetry energy curvature coefficient for Ne, Na, Mg, Al, and Si nuclei from the proton to neutron drip-lines. The Coherent Density Fluctuation Model (CDFM) is used to estimate these quantities taking the relativistic mean-field densities as inputs. The Br\"uckner energy density functional is taken for the nuclear matter binding energy and local density approximation is applied for its conversion to coordinate space. The symmetry energy again decomposed to the volume and surface components within the liquid drop model formalism to the volume and surface parts separately. Before calculating the surface properties of finite nuclei, the calculated bulk properties are compared with the experimental data, whenever available. The NL3* parameter set with the BCS pairing approach in an axially deformed frame-work is used to take care of the pairing correlation when needed. The deformed density is converted to its spherical equivalent with a two Gaussian fitting, which is used as an input for the calculation of weight function in the CDFM approximation. With the help of the symmetric energy for the recently isotopes $^{29}$F, $^{28}$Ne, $^{29,30}$Na and $^{31,35,36}$Mg are considered to be within the {\it  island of inversion} emphasized {\bf [Phys. Lett. B 772, 529 (2017)]}. Although we get large symmetric energies corresponding to a few neutron numbers for this isotopic chain as expected, an irregular trend appears for all these considered nuclei. The possible reason behind this abnormal behavior of symmetry energy for these lighter mass nuclei is also included in the discussion, which gives a direction for future analysis.
\end{abstract}
\maketitle
%%%%%%%%%%%%%%%%
\section{Introduction}
\label{sec1}
The surface property of a nucleus is one of the pivotal quantities to determine the structural information of finite and infinite nuclear matter at isospin asymmetry. The symmetric $S$ energy determines the nuclear matter equation of state (EoS), which is connected with the isospin asymmetry. This $S$ controls the properties of the neutron star (NS), such as its mass and radius \cite{li2008}. Consequently, the Gravitational Waves (GW), the parameter of NS merger, etc. are automatically governed by the surface properties. Because of the interrelation of the nuclear matter symmetric energy with the neutron-skin thickness of finite nuclei, one can study the NS structure knowing the symmetric energy of the nuclear matter EoS \cite{li2019}. Similarly, there is a strong bond between the nuclear matter surface properties with the corresponding quantities of finite nuclei. There has been many significant  efforts both in nuclear physics and astrophysics  over the last two decades, and much progress in this regard is constraining both the magnitude of the symmetry energy $S(\rho_0)$ and the slope parameter $L_{sym}^{A}$ = $3\rho_0 (dS(\rho)/d\rho)_{\rho_0}$  at the saturation density $\rho_0$ of nuclear matter \cite{baran2005,li2008,tsang2012,horowitz2014,baldo2016,xu2020}. \\
%%%%

Recently, a large number of works have been reported, which connect the surface properties of finite nuclei with the nuclear matter observables \cite{anto18,anto2,anto3,anto4,gad11,gad12,bhu18,bhu20}. To calculate these properties, a large number of methods are adopted such as Coherent Density Fluctuation Model \cite{anto1,anto2,anto3,anto4,gad12,bhu18} and Liquid Drop Approximation \cite{dani03}. Recently, the Coherent Density Fluctuation Model (CDFM) is gaining momentum for the evaluation of such properties. Although the original idea is published \cite{antozphys1980} long back, substantial work is waiting yet. Some of the relations of symmetric energy and related quantities have a strong correlation with the structures of the finite nuclei. We have reported in an isotopic series that the symmetric energy shows a peak at the corresponding magic number \cite{bhu18,bhu20,manpreet}. Due to the advancement of the radioactive ion beam facility, the study of highly asymmetric isospin neutron/proton-rich nuclei open up a new area of research in Nuclear Physics. Some of the neutron number for certain proton combination behaves like a magic number. For example, neutron number N= 14, 16 and  32/34 act like close shell combination for neutron-rich Ne and Mg isotopes \cite{tanihata}. The nuclei Ne, Na, Mg, Al, and Si show unexpected characteristics near the neutron-rich region. Some of the isotopic region termed as {\it  island of inversion}. In this region of the mass table, the experimental binding energy deviates considerably from the shell model calculations. Long back Patra and Praharaj \cite{patra91} shown that these isotopes exhibit a large deformation in their ground states, which is later on verified by Tanihata {\it  et al.} \cite{tanihata} in their experimental study. In this paper, we would like to perform a systematic study of surface properties of these nuclei starting from the proton deficient region to the neutron-drip line and look for various structures. \\
%%%%%

A large number of theoretical formalisms are available in the literature for the structural analysis of finite nuclei and EoS of infinite nuclear matter. Among those well-known approaches, the Skyrme Hartree-Fock and relativistic mean-field (RMF) theory are the most successful frame-work for such study. The RMF approach has an advantage over the non-relativistic counterpart because of the inclusion of the spin-orbit interaction through the relativistic equations. This approach also reproduces the binding energy, quadruple deformation parameter $\beta_2$, and other related properties of finite nuclei throughout the mass table. With a few parameters of meson-nucleon couplings and their respective masses, the method successfully reproduces the experimental quantities for almost all known nuclei. This method also explains the nuclear EoS and NS properties quite well. Because of the satisfactory reproduction of experimental quantities by using RMF formalism, in the present paper, we use the formalism to get the bulk properties of finite nuclei including densities. These densities are used in the CDFM to generate surface properties of finite nuclei.
%%%%%

The paper is arranged as follows: The relativistic mean-field model and the fitting procedure of energy density functional ${\cal{E}(\rho)}$ to an analytical expression in coordinate space with the help of the Br\"uckner prescription \cite{bruk68,bruk69} is discussed in Sec. \ref{theory}. Also, the Coherent Density Fluctuation Model is briefly outlined. The Sec. \ref{results} is dedicated to the discussion of our results derived from the calculations. A summary and concluding remarks are presented in Sec. \ref{summary}.
%%%%%%

%
\section{Relativistic mean field (RMF) Model}
%\label{theory}
%
%\subsection{E-RMF Formalism}\label{ERMF}
\label{theory}
In this section, we briefly describe the formalism of the well documented relativistic mean-field theory. The standard nonlinear RMF Lagrangian density is built up by the interactions of isoscalar-scalar $\sigma$, isoscalar-vector $\omega$, and isovector-vector $\rho$ mesons with nucleons. In this version of the Lagrangian, the self-interaction of the $\sigma-$meson is also considered. Recently, the extended versions of the relativistic mean-field formalism based on the effective field theory motivated relativistic mean-field formalism are very popular \cite{frun96,frun97}. In these models, all possible meson-nucleon and their self-interactions are considered. The standard RMF formalism is quite successful for finite nuclei not only for $\beta-$stable but also predicts reasonably the properties of drip-lines and super-heavy nuclei \cite{kumar18,kumar17}. Thus, we have considered this model in the present work. The RMF Lagrangian is discussed in detail in Refs. \cite{kumar18,kumar17,frun96,frun97}. During the last few decades, the pertinence of this decorum to nuclear astrophysics is in limelight. It foresees the structure of NS and reveals the tidal deformability adequately \cite{malik2018}. The recently reported experimental data of Gravitational Wave \cite{GW170817} measurements constraint the RMF nuclear EoS in a precise manner. Thus, the RMF Lagrangian gives enough confidence to adopt for further implementations to different nuclear environments. For a nucleon-meson interacting system the energy density functional is granted as \cite{kumar18}:
%%%%%%
\begin{widetext}
\begin{eqnarray}
{\cal E}({r})&=&\sum_{\alpha=p,n} \varphi_\alpha^\dagger({r})\Bigg\{-i \mbox{\boldmath$\alpha$} \!\cdot\!\mbox{\boldmath$\nabla$}+\beta \bigg[M-\Phi (r)\bigg]+ W({r})+\frac{1}{2}\tau_3 R({r})+\frac{1+\tau_3}{2} A({r})
\bigg)\Bigg\} \varphi_\alpha(r)
\nonumber\\
&&\
+\left(\frac{1}{2}+\frac{\kappa_3}{3!}\frac{\Phi({r})}{M}+\frac{\kappa_4}{4!}\frac{\Phi^2({r})}{M^2}\right)\frac{m_s^2}{g_s^2}\Phi^2({r})
-\frac{1}{2}\frac{m_\omega^2}{g_\omega^2} W^2({r})
-\frac{1}{2}\frac{m_\rho^2}{g_\rho^2} R^2({r})\;,
\label{edf}
\end{eqnarray}
\end{widetext}
%%%%%%
Here, $\Phi$, $W$ and $R$ are the re-considered fields for $\sigma$, $\omega$ and  $\rho$ mesons written as  $\Phi = g_s\sigma_{0} $, $W = g_\omega \omega_{0}$ and  $R = g_\rho\rho $ respectively. $M$, $m_{\sigma}$, $m_{\omega}$ and  $m_{\rho}$ are the nucleon masses $\omega$ , $\sigma$ and $\rho$  mesons, sequentially. From Eq. (\ref{edf}), we obtain our energy density ${\cal{E}}_{nucl.}$ \cite{kumar18,kumar17} by taking that the exchange of mesons establish an uniform field, where the oscillations done by nucleons in a periodic motion said to be simple harmonic . From the effective-RMF energy density, the equation of motions (EoS) for the mesons and the nucleons are procured by using the Euler-Lagrange equation. A bunch of coupled differential equations are retained and settled accordingly \cite{kumar18}. The scalar and vector densities,
%%%%%
\begin{eqnarray}
\rho_s(r)&=&\sum_\alpha \varphi_\alpha^\dagger({r})\beta\varphi_\alpha, \label{scaden}\\
\rho_v(r)&=&\sum_\alpha \varphi_\alpha^\dagger({r})\tau_{3}\varphi_\alpha\label{vecden},
\end{eqnarray}
%%%%%
are figured out from the converged elucidations within the spherical harmonics.
%%%%%

\subsection{Spherical Equivalent Density using two Gaussian Fitting }
We have obtained the axially deformed density from RMF model for considered parameter sets and converted it into it's spherical equivalent density by following the below steps,
\begin{itemize}
\item Initially we have the deformed density $\rho(r_{\bot},z)$ which is converted to one dimensional as $\bar{\rho}(\omega)$
where $\omega =\sqrt{x^{2}+y^{2}}$ by performing the z-integration over the whole space as done in \cite{patra2009,panda2014,sharma2016}:
\begin{equation} %\label{eq4}
 \vec{\rho}(\omega) = \int_{-\infty}^{\infty} \rho(\sqrt{\omega^2+z^2}) dz \;.
 \label{cyld}
\end{equation} 
\item In the second step, the z-integrated density $\bar{\rho}(\omega)$ is fitted to a two Gaussian function expressed as:
\begin{equation} %\label{eq5}
\rho(r) = \sum_{i=1}^2 c_i exp[-a_ir^2] \;,
\label{eqvd}
\end{equation}
where the co-efficient $c_{i}$ and range $a_{i}$ are given with initial values for respective nuclei. 
\end{itemize}
 Then we have used the spherical equivalent density in place of deformed density for further calculations. The spherical equivalent densities are normalized to mass number of the nucleus. \\
%%%%%

To carry a detailed study of the nuclear bulk properties of open-shell nuclei, the contribution of pairing correlation becomes a major factor. In the present calculations, we will be using BCS approach for the calculation of pairing correlation. Here, the RMF+BCS formalism with NL3* parameter set \cite{lala09} is used for the finite nuclei. Within this approach, we evaluate the binding energy, quadruple deformation parameter $\beta_2$, root mean square radii and the axially deformed densities for both neutrons and protons distribution. The vector density $\rho_v(r)$ is farther handled within the CDFM to locate the weight function $|F(x)|^2$, which is an extensive quantity to reckon the symmetry energy ($S^{A}$), neutron pressure ($P^{A}$) and symmetry energy curvature coefficient ($K_{sym}^{A}$) for Ne, Na, Mg, Al and Si nuclei in the isotopic chains. \\
%%%%%%%%%%%%%%%%%%%%%%%%%%%%%%%%%%%%%%%%

\subsection{Nuclear Matter Parameter}\label{fitting}
The expression for the energy density of infinite and isotropic nuclear matter (NM) are obtained from the Br\"{u}ckner functional defined as \cite{bruk68,bruk69}:
%%%
\begin{eqnarray}
{\cal{E}}(\rho)_{nucl.}&=&AV_0(x)+V_C-V_{Cx},
\label{brue0}
\end{eqnarray}
where
%%%%
\begin{eqnarray}
V_0(x)&=& 37.53\Big[(1+\delta)^{5/3} + (1-\delta)^{5/3}]\rho_0(x)^{2/3} \nonumber \\ 
&& + b_1\rho_0(x) +b_2\rho_0(x)^{4/3} + b_3\rho_0(x)^{5/3} \nonumber \\
&& +\delta^2[b_4\rho_0(x) + b_5\rho_0(x)^{4/3} + b_6 \rho_0(x)^{5/3} \Big].
\end{eqnarray}
%%%%
Here, $b_1= -741.28$, $b_2=1179.89$, $b_3=-467.54$, $b_4=148.26$,
$b_5=372.84$, $b_6=-769.57$. In each {\it  Flucton} there are protons having Coulomb energy $V_C=\frac{3}{5}\frac{Z^2e^2}{x}$
and Coulomb exchange energy $V_{Cx}=0.7386Ze^2(3Z/4\pi x^3)^{1/3}$.
%%%%%

The important part of the present calculation is to convert the NM quantities Eq. (\ref{brue0}) from momentum ($\rho-$) space to coordinate ($r-$) space in local density approximation (LDA). The NM parameters $S^{NM}$, $P^{NM}$, $L_{sym}^{NM}$ and $K_{sym}^{NM}$ are obtained from the well defined relations \cite{kumar18,Chen2014,bruk68}:
%%%%%%
\begin{eqnarray}
S^{NM}&=&\frac{1}{2}\frac{\partial^2 ({\cal E}/\rho)}{\partial\alpha^2}\Big|_{\alpha=0},\label{snm1}\\
L_{sym}^{NM}&=&3\rho\frac{\partial S(\rho)}{\partial\rho}\Big|_{\rho=\rho_0} = \frac{3P^{NM}}{\rho_{0}},\label{lsymnm}\\
K_{sym}^{NM}&=&9\rho^2\frac{\partial^2 S(\rho)}{\partial\rho^2}\Big|_{\rho=\rho_0}.\label{ksymnm}
\end{eqnarray}
%%%%%%
Here, 
\begin{eqnarray}
S^{NM} &=& 41.7\rho_0(x)^{2/3}+b_4\rho_0(x)+b_5\rho_0(x)^{4/3}
\nonumber\\
&& + b_6\rho_0(x)^{5/3},\label{snm} \\ 
P^{NM} &=& 27.8\rho_0(x)^{5/3}+b_4\rho_0(x)^2+\frac4 3 b_5\rho_0(x)^{7/3} \nonumber\\
&& +\frac5 3 b_6\rho_0(x)^{8/3}, \label{pnm} \\
K_{sym}^{NM} &=& -83.4\rho_0(x)^{2/3}+4b_5\rho_0(x)^{4/3} \nonumber\\
&& +10b_6\rho_0(x)^{5/3},  
\label{ksymn}
\end{eqnarray}
are the nuclear matter quantities at local density. The nuclear densities of Ne, Na, Mg, Al and Si nuclei are calculated using RMF formalism. These densities are used as input in CDFM (described in the following sub-section) to calculate our weight function, which is a key quantity acting as a overpass among the NM parameters within ($\rho-$) space and  in ($r-$) space of finite nuclei (manipulating LDA). To contest with the coordinate ($r-$) and momentum ($\rho-$) space simultaneously, we can build up the total density of a nucleus. The number of Fluctons in an infinite manner can be superimposed to form the total density of the nucleus, following the approach of CDFM discussed below.
%%%%%%

\subsection{Coherent Density Fluctuation Model}\label{CDFM}
The Coherent Density Fluctuations Model (CDFM) is firstly prescribed by Antonov et al \cite{antozphys1980}, which plays a vital role to take care of the fluctuation of momentum and coordinate. This method can be easily used to interpret the surface properties of finite nuclei. In this CDFM formalism we can use NM tools $S^{NM}$, $P^{NM}$ and $K_{sym}^{NM}$ from Eqs. (\ref{snm1})$-$(\ref{ksymn}) to obtain their values for a finite nucleus \cite{anto1,anto2,anto3,antozphys1980}. 
Within this model, the density  $\rho$ ({\bf  r, r$'$}) of a finite nucleus can be rewritten as the coherent superposition of infinite number of one-body density matrix (OBDM)  $\rho_x$ ({\bf  r}, {\bf  r$'$}) for spherical parts of NM  coined as {\it  Fluctons} \cite{bhu18,gad11},
\begin{equation}
\rho_x ({\bf  r}) = \rho_0 (x)\, \Theta (x - \vert {\bf  r} \vert),
\label{denx}
\end{equation}
where $\rho_o (x) = \frac{3A}{4 \pi x^3}$. The generator coordinate x is the radius of a sphere consisting of Fermi gas having all the  A nucleons distributed uniformly over it. It is appropriate to administer for such a system the OBDM disclosed as below \cite{bhu18,anto2,gad11,gad12},
\begin{equation}
\rho ({\bf  r}, {\bf  r'}) = \int_0^{\infty} dx \vert F(x) \vert^2 \rho_x ({\bf  r}, {\bf  r'}),
\label{denr}
\end{equation}
where $\vert F(x) \vert^2 $ is called as weight function (WF). The coherent superposition of OBDM $\rho_x ({\bf  r}, {\bf  r'})$ is given below  as:
\begin{eqnarray}
\rho_x ({\bf  r}, {\bf  r'}) &=& 3 \rho_0 (x) \frac{J_1 \left( k_f (x) \vert {\bf  r} - {\bf  r'} \vert \right)}{\left( k_f (x) \vert {\bf  r} - {\bf  r'} \vert \right)} \nonumber \\
&&\times \Theta \left(x-\frac{ \vert {\bf  r} + {\bf  r'} \vert }{2} \right),
\label{denrr}
\end{eqnarray}
where J$_1$ is said to be the spherical Bessel function kind of first order and $k_{f}$ is the Fermi momentum of nucleons inside the {\it  Flucton} having radius $x$ and  $k_f (x)=(3\pi^2/2\rho_0(x))^{1/3} =\gamma/x$, ($\gamma\approx 1.52A^{1/3}$). The Wigner distribution function of the OBDM of Eq. (\ref{denrr}) is given by,
\begin{eqnarray}
W ({\bf  r}, {\bf  k}) =  \int_0^{\infty} dx\, \vert F(x) \vert^2\, W_x ({\bf  r}, {\bf  k}).
\label{wing}
\end{eqnarray}
Here, $W_x ({\bf  r}, {\bf  k})=\frac{4}{8\pi^3}\Theta (x-\vert {\bf  r} \vert)\Theta (k_F(x)-\vert {\bf  k} \vert)$. The density $\rho$ (r) in terms of weight function within the CDFM access is:
\begin{eqnarray}
\rho (r) &=& \int d{\bf  k} W ({\bf  r}, {\bf  k}) \nonumber \\
&& = \int_0^{\infty} dx\, \vert F(x) \vert^2\, \frac{3A}{4\pi x^3} \Theta(x-\vert{\bf  r} \vert),
\label{rhor}
\end{eqnarray}
%%%%%%
which is normalized to A, i.e.,  $\int \rho ({\bf  r})d{\bf  r} = A$. In the $\delta$-function limit, the Hill-Wheeler integral equation, that is the differential equation for the WF in the generator coordinate is retrieved \cite{anto1}. The $|F(x)|^2$ for a provided density $\rho$ (r) is described as here
%%%%%%
\begin{equation}
|F(x)|^2 = - \left (\frac{1}{\rho_0 (x)} \frac{d\rho (r)}{dr}\right)_{r=x},
\label{wfn}
\end{equation}
%%%%%%
with $\int_0^{\infty} dx \vert F(x) \vert^2 =1$. A detailed genealogy can be found in Refs. \cite{bhu18,anto1,anto2,gad11,gad12}. The finite nuclear symmetry energy $S^{A}$ neutron pressure $P^{A}$ and surface curvature coefficients $K_{sym}^{A}$ are calculated by weighting the corresponding quantities for infinite NM within the CDFM, as given below \cite{anto4,gad11,gad12,fuch95,anto17}
%%%%%%
\begin{eqnarray}
S^{A}= \int_0^{\infty} dx\, \vert F(x) \vert^2\, S^{NM} (\rho (x)) ,
\label{s0}
\end{eqnarray}
\begin{eqnarray}
P^{A} =  \int_0^{\infty} dx\, \vert F(x) \vert^2\, P^{NM} (\rho (x)),
\label{p0}
\end{eqnarray}
\begin{eqnarray}
K_{sym}^{A} =  \int_0^{\infty} dx\, \vert F(x) \vert^2 \ K_{sym}^{NM} (\rho (x)).
\label{k0}
\end{eqnarray}
%%%%%%
The $S^{A}$, $P^{A}$ and $K_{sym}^{A}$ in  Eqs. ($\ref{s0}-$\ref{k0}) are the surface weighted average of the corresponding NM quantities in the LDA limit for finite nuclei.

\subsection{Volume and surface symmetric energy in Danielewicz's liquid drop prescription}
The nuclear binding energy E(A, Z) in the liquid drop model incorporating the volume symmetry energy parameter $S_V$ and modified surface symmetry energy parameter $S_S$ is written as \cite{steiner2005, myers}:
\begin{eqnarray}
E(A,Z) = -B.A+E_{S}A^{2/3}+S_{V}A \frac{(1-2Z/A)^{2}}{1+S_{S}A^{-1/3}/S_{V}} \nonumber \\
+E_{C}\frac{Z^{2}}{A^{1/3}}+E_{dif}\frac{Z^{2}}{A}+E_{ex}\frac{Z^{4/3}}{A^{1/3}}+a \Delta A^{-1/2}.
\label{eq3}
\end{eqnarray}
In Eq. (\ref{eq3}), B= binding energy per particle of symmetric NM at saturation. $E_{S}$, $E_{C}$, $E_{dif}$, and $E_{ex}$ are the coefficients of the surface energy of symmetric matter, the Coulomb energy, the diffuseness correction, and the exchange correction to the Coulomb energy, conjointly. The pairing corrections are delivered by the uttermost term, which is crucial for open-shell nuclei. The symmetry energy is rewritten as [3rd term of Eq. (\ref{eq3}) in the form S = $(N-Z)^{2}$/A, where
\begin{eqnarray}
S= \frac{S_{V}}{1+ \frac {S_{S}} {S_{V}} A^{-1/3}}= \frac{S_{V}}{1+A^{-1/3}/\kappa}.
\label{eq4}
\end{eqnarray}
From the above Eq. (\ref{eq4}), the individual components of $S_V$ and $S_S$ can be written as:
\begin{eqnarray}
S_{V}=  S \left (1+\frac{1}{\kappa A^{1/3}} \right)
\end{eqnarray}
and
\begin{eqnarray}
S_{S}= \frac {S}{\kappa} \left (1+\frac{1}{\kappa A^{1/3}} \right).
\end{eqnarray}
%%%%%%%%%%

The symmetry energy and its volume and surface components are calculated within the  CDFM formalism \cite{bhu18,anto2,antozphys1980}. Following Refs. \cite{dani03,dani04,dani06}, an approximate expression for the ratio $\kappa \equiv \frac{S^{V}}{S^{S}}$ can be written within the CDFM,
\begin{eqnarray}
\kappa = \frac{3}{R\rho_{0}}\int_0^{\infty} dx \vert F(x) \vert^2 x \rho_{0}(x) \left [\left(\frac{S(\rho_{0})}{S(\rho(x)}\right )-1 \right].
\label{kappa}
\end{eqnarray}
Here S($\rho_{0}$) is the nuclear symmetry energy at equilibrium nuclear matter density $\rho_{0}$. Employing the density dependence of symmetry energy \cite{dani03}:
\begin{eqnarray}
S[\rho(x)] = S^{V} \left(\frac{\rho(x)}{\rho_{0}}\right)^{\gamma} ,
\label{s0x}
\end{eqnarray}
There exist various estimations for the value of the parameter $\gamma$. In present work, we use $\gamma$ = 0.3 in reference to \cite{anto18}. Using above eq. and S($\rho_{0}$) = $S_{V}$, the Eqs. (\ref{s0}) and (\ref{kappa}) can be re-written as follows:
\begin{eqnarray}
S = S(\rho_0)\int_0^{\infty} dx \vert F(x) \vert^2 \left (\frac{\rho(x)}
{\rho_{0}}\right )^{\gamma},
\label{s0t}
\end{eqnarray} 
and
\begin{eqnarray}
\kappa = \frac{3}{R\rho_{0}}\int_0^{\infty} dx \vert F(x) \vert^2 x \rho_{0}(x)
\left (\left (\frac{\rho_{0}}{\rho(x)}\right )^{\gamma}-1\right).
\label{k0t}
\end{eqnarray}
%%
%%%%%%%%%%%%%%%%%%%%%%%%%%%%%%
\begin{figure}
\includegraphics[width=1.05\columnwidth]{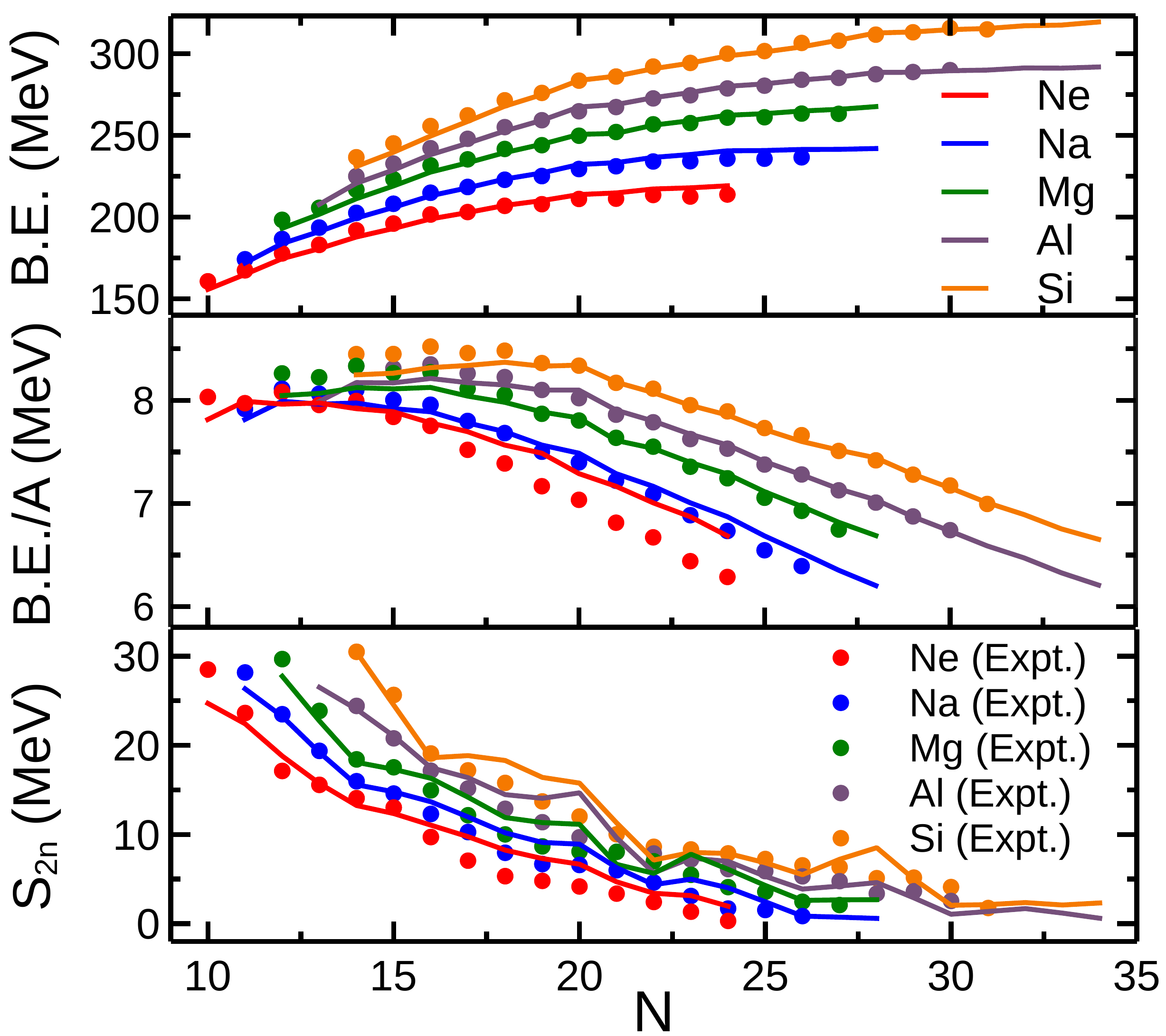}
\caption{(Color online)  (a) The total binding energy B.E. (b) Binding energy per particle B.E./A and  (c) Two neutron separation energy $S_{2n}$ as a function of neutron number N for Ne, Na, Mg, Al and Si isotopes, respectively.}
\label{NMEOS}
\end{figure}
%%%%%%%%%%%%%%%%%%%%%%
\begin{figure}
\centering
\includegraphics[width=1.05\columnwidth]
{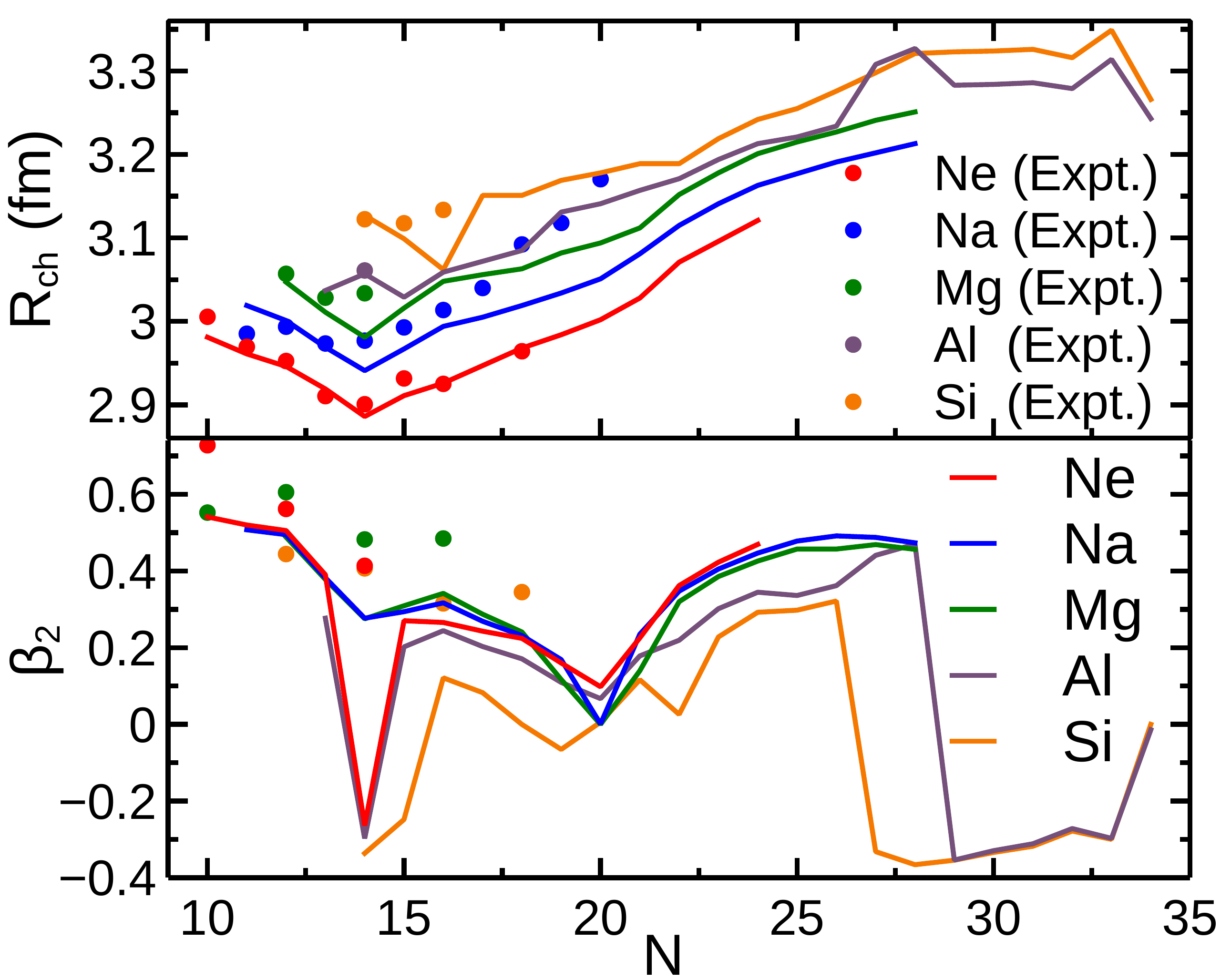}
\caption{(color online) (a) The charge distribution radius and (b) the quadruple deformation parameter $\beta_2$ as a function of neutron number N for Ne, Na, Mg, Al, and Si isotopes.}
\label{BE}
\end{figure}
%%%%%%%%%%%%%%%%%%%%%%%%%%%%%%
\section{Results and discussions}
\label{results}
\begin{figure}[H]
\centering
\includegraphics[width=1\columnwidth]{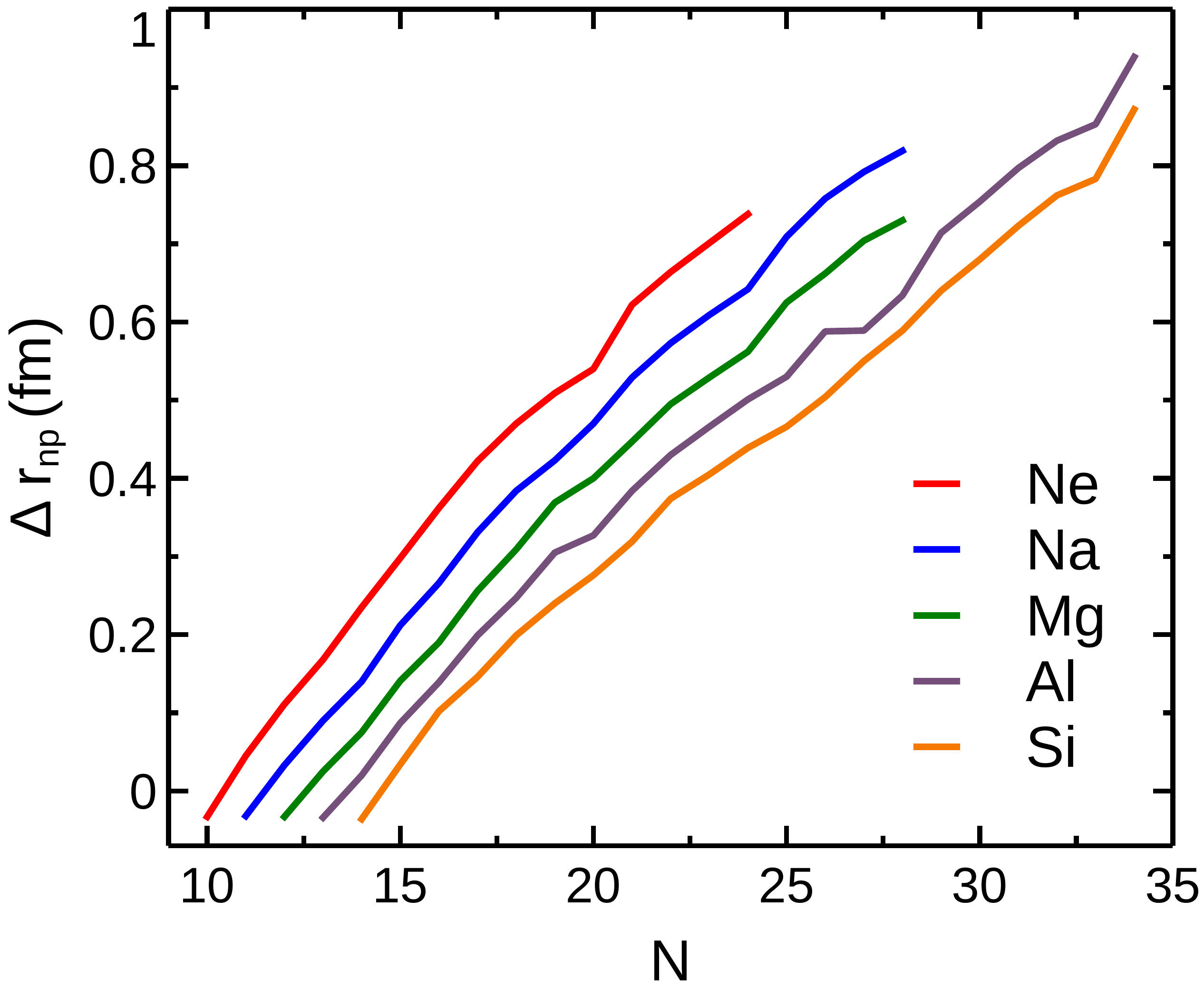}
\caption{(color online) The neutron skin thickness $\triangle{r_{np}}$ is presented as a function of neutron number N for Ne, Na, Mg, Al and Si nuclei.}
\label{nsthickness}
\end{figure}

\begin{figure}[H]
\centering
\includegraphics[width=0.5\textwidth]{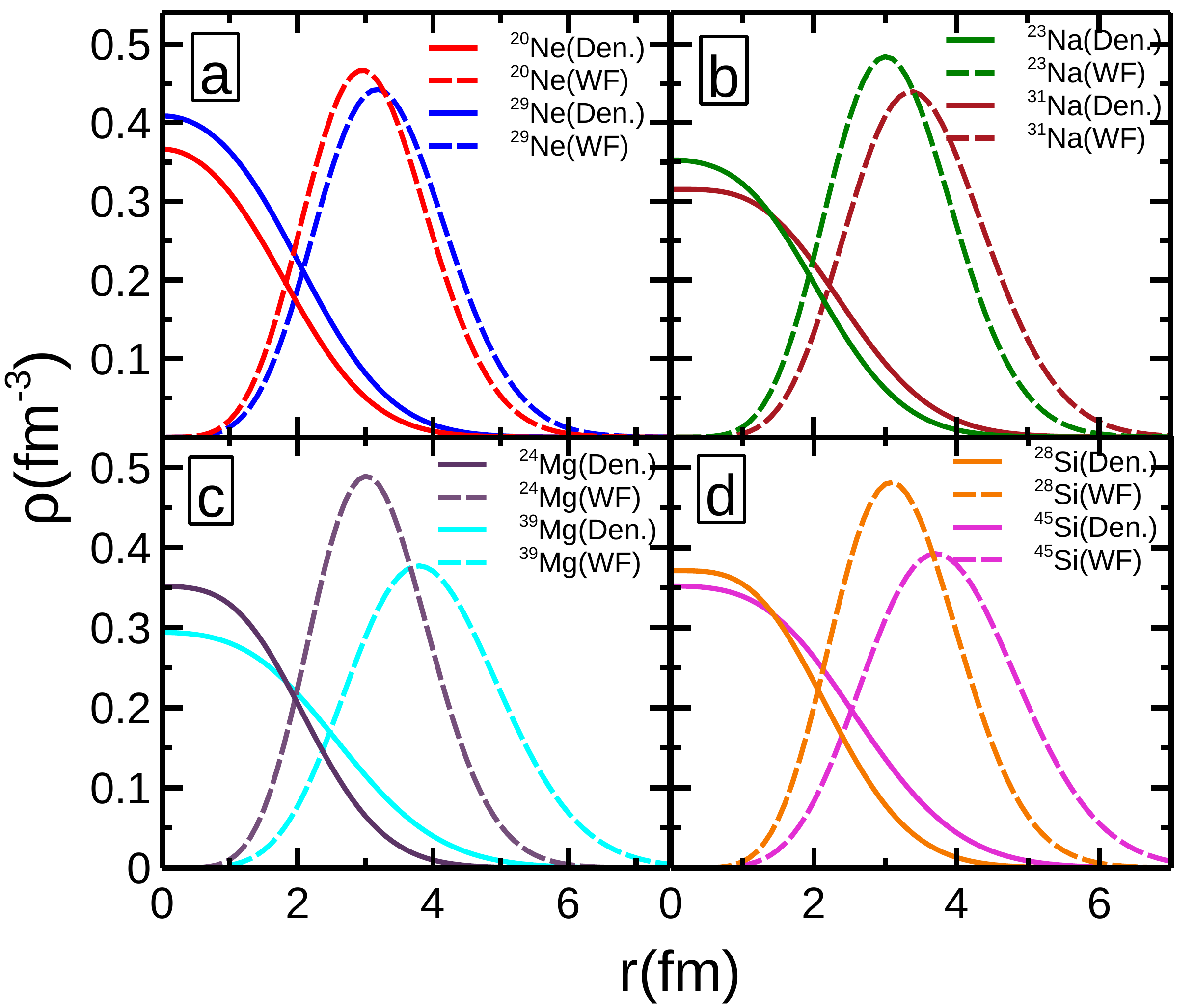}
\caption{(color online) The spherical equivalent density (solid line) for some of the selected nuclei are shown. The corresponding weight function (dashed line) are also displayed.}
\label{diffuse}
\end{figure}

The surface properties, like symmetric energy, neutron pressure, and symmetry energy curvature coefficient of finite nuclei are studied using the coherent density fluctuation model by taking the RMF density as input in the calculations. Before going to the explanation of the surface properties of finite nuclei, we highlight herewith the bulk properties (binding energy, deformation, and nuclear charge radius) of the considered nuclei such as Ne, Na, Mg, Al, and Si. The results are discussed through Table \ref{tab1} and Figs. \ref{NMEOS}-\ref{eosden}. In Table \ref{tab1} , the values of the coefficients of the Gaussian function [Eq. \ref{eqvd}] $c_1$, $c_2$ and $a_1$, $a_2$ are given, which are used to get the spherical equivalent densities of the Ne, Na, Mg, Al and Si isotopes. The obtained spherical equivalent densities are further used to calculate the weight function $|F(x)|^2$ [Eq. \ref{k0}]. Then the weight function $|F(x)|^2$ is used to calculate the nuclear matter parameters, such as symmetric energy  $S^A$, neutron pressure $P^A$ and symmetry energy curvature coefficient $K_{sym}^A$ for finite nuclei. The predicted results are given in Figs. \ref{NMEOS}-\ref{eosden} in the following subsections.

\subsection{Binding energy and two neutron-separation energy} To estimate the surface properties such as symmetry energy, neutron pressure, and symmetry energy curvature coefficient of finite nuclei, one needs to have an understanding of the ground state bulk properties. The binding energy (BE), root mean square (rms), charge radius ($R_{ch}$) and quadruple deformation parameter ($\beta_2$) for Ne, Na, Mg, Al, and Si isotopes starting from proton-rich to the expected neutron drip-line are obtained by using RMF model and NL3$^*$ parameter set. The calculated results with available experimental data for the above-considered isotopes are shown in Figures \ref{NMEOS} and \ref{BE}. From the overall observation of binding energy for all the considered isotopic chains, we note a good agreement between our results and the experimental data \cite{raman,angeli}. A careful analysis of Fig. \ref{NMEOS} shows that the expected drip-lines for Ne, Na, Mg, Al, and Si isotopic chain are at mass number (A) = 34, 39, 40, 47, and 48 respectively.
%%%%%%%%%%%%%%%%%%%%%%%%%%%%%%
%%%%%%%%%%%%%%%%%%%%
%%%%%%%%%%%%%%%%%%%%%%%%%%%%%%
The two neutron separation energy $S_{2n}$ can be determined using the relation $S_{2n} (N,Z)$= BE(N,Z) - BE(N-2,Z). Where both the $BE's$ are evaluated using RMF(NL3*) parameter for theory and experimental values are taken from the Ref. \cite{wang2017} for comparison. The estimated $S_{2n}$ values along with the available experimental data are presented in the lower panel of Fig. \ref{NMEOS} for the $^{18-34}$Ne, $^{20-39}$Na, $^{22-40}$Mg, $^{24-47}$Al and $^{26-48}$Si isotopic series. One can notice from the figure, $S_{2n}$ decreases smoothly with an increase in neutron number except some sharp discontinuities (i.e kinks) at some particular neutron number $N_{magic}$ indicating the shell closure property. It is worthy here to mention that the energy requires to remove two neutrons from a nucleus with $N_{magic}$+2 is much less than the nucleus with $N_{magic}$ breaking the regular pattern of the $S_{2n}$. In the lower panel of Fig. \ref{BE}, we have displayed the total quadruple deformation parameter $\beta_2$ for the considered isotopic series. It is clear from the figure that there are shape transitions at neutron numbers (N) = 14 for Ne, 20 for Na and Mg, 28 for Si isotopic series.
%%%%

To find the ground state solution of a given nucleus, we obtain the conversed solutions by taking various initial quadruple deformation. From all those solutions, we have considered the maximum binding energy and corresponding quantities as the ground state of the given nucleus. The radial components, which are having odd under time-reversal symmetry and parity must be neglected. In the present paper, we have implemented the blocking method which rectifies the time-reversal symmetry. Normally we need to block different states around the Fermi level to determine the lowest energy states for the odd nucleus as per odd-odd, odd-even, and even-odd nuclei, the time-reversal symmetry violates. In order to take care of the time-reversal violation into account, we used the simple blocking procedure as discussed in Ref. \cite{patra}. In many of the cases, we get almost nearly equal energies among various states indicating the shape co-existence nature of the isotopes. As we are following the criteria of taking the highest BE result, there are some low-laying excited states (degenerate states), which may be the reason behind the irregular plots. From those obtained values we got our ground state, excited states and degenerate states (almost same magnitude of binding energy with the different state).  

In Fig. \ref{BE}, we have displayed the charge distribution radius $R_{ch}$ (upper panel) and quadruple deformation parameter $\beta_2$ (lower panel) for the isotopic chain of Ne, Na, Mg, Al, and Si nuclei. The experimental data \cite{raman,angeli} are also compared, wherever available. Here in the lower mass region, the charge radius $R_{ch}$ decreases and then increases with the mass number for all the isotopic chain. From the quadruple deformation parameter ($\beta_2$), we find shape transition appears at N = 14, 20, 20, 29, and 28 for Ne, Na, Mg, Al, and Si isotopes, respectively. It is to be noted that the irregular nature of the results mostly due to the shape co-existence nature of the isotopes. A low-lying excited state is very often noticed in these isotopic chains.
%%%%%

\subsection{The neutron skin thickness} %$\triangle{r_{np}}=r_n-r_p$}
In Fig. \ref{nsthickness}, we have presented the neutron skin thickness $\triangle r_{np}= r_n – r_p$ as a function of neutron number. The $\triangle r_{np}$ is an important quantity, which is connected with the surface property of the nucleus in terms of isospin asymmetry. It has a direct relation with the nuclear equation of state (EoS) which controls the structure of the neutron star. Thus, it has an important role in nuclear astrophysics. We have shown it for Ne, Na, Mg, Al, and Si isotopic series. As we see the neutron skin thickness increases with neutron number, i.e., the presence of more neutrons enhanced the nuclear radius of the nucleus. The neutron skin thickness in the nucleus is formed by a combination of volume and surface contributions. The volume part explains an increase in the local mean field of the surface of the neutron to the proton individual and the surface part in which the surface width of the neutron increases as compared to one of the individual protons. Both the contributions increase with neutron excess \cite{vinas2012} in an isotopic series. The linear increase in neutron skin thickness indicates the surface/volume saturation, which also tells the increase in symmetric energy.

%%%%%%%%%%%%%%%%%%%%%%%%%%%%%
\subsection{The spherical equivalent density and weight function }
\label{equiv}
In Fig. \ref{diffuse}, we have plotted the spherical equivalent densities (solid) and corresponding weight functions $\vert F(x) \vert^2 $ (dashed) as a function of radius. We have converted the deformed densities into their spherical equivalents through two Gaussian fittings by using Eq. \ref{eqvd}. The spherical equivalent density for any nucleus can be easily obtained by using the values of $c_1$, $a_1$ and $c_2$, $a_2$ given in Table \ref{tab1}. For a careful inspection of Table \ref{tab1}, it is noticed that for all the nuclei, the values for $c_1$ are found to be negative and $c_2$ with a positive values. However, the values of $a_1$ and $a_2$ are almost comparable to each other with positive signs. More detail analysis of the fitting and also technical details can be found in Refs. \cite{patra2009,panda2014,sharma2016}. 
\begin{table*}
\caption{The axially deformed RMF density obtained from the NL3$^*$ parameter set are converted to its spherical equivalent using two Gaussian functionals. The coefficients $c_1$, $a_1$ and $c_2$, $a_2$ are listed for Ne, Na, Mg, Al, and Si isotopes.}
\renewcommand{\tabcolsep}{0.22cm}
\renewcommand{\arraystretch}{1.0}
\begin{tabular}{cccccccccc}
\hline\hline 
Nucleus & $c_1$& $a_1$& $c_2$ &$a_2$ & 
Nucleus & $c_1$& $a_1$& $c_2$ &$a_2$\\
\hline
$^{19}$Ne & -3.16481 &0.331586  &3.47687  &0.311655 & 
$^{20}$Ne & -3.04522 &0.333589  &3.41182  &0.313962 \\ 
$^{21}$Ne & -3.59429 &0.337037  &3.95496  &0.317189 &
$^{22}$Ne & -4.08107 &0.339088  &4.4364   &0.319048 \\
$^{23}$Ne & -3.83833 &0.316714  &4.17055  &0.297795 &
$^{24}$Ne & -3.61795 &0.42193   &3.99949  &0.373673 \\
$^{25}$Ne & -1.61846 &0.409952  &2.0865   &0.337241 &
$^{26}$Ne & -1.44537 &0.400685  &1.90112  &0.322851 \\
$^{27}$Ne & -0.856237&0.394508  &1.30403  &0.291229 &
$^{28}$Ne & -0.805704&0.365513  &1.24565  &0.274331 \\
$^{29}$Ne & -2.09689 &0.306084  &2.50567  &0.272813 &
$^{30}$Ne & -3.85888 &0.276814  &4.23455  &0.260330 \\
$^{31}$Ne & -2.18123 &0.208782  &2.44085  &0.194627 &
$^{32}$Ne & -1.55223 &0.202165  &1.8495   &0.187297 \\
$^{33}$Ne & -1.07001 &0.210559  &1.37341  &0.186965 &
$^{34}$Ne & -0.284251&0.253762  &0.588779 &0.171028 \\
\hline
$^{20}$Na & -3.30838 &0.33129   &3.61359  &0.309985 &
$^{21}$Na & -3.36721 &0.336049  &3.72673  &0.315269 \\
$^{22}$Na & -4.20169 &0.339772  &4.55721  &0.320112 &
$^{23}$Na & -4.93147 &0.342127  &5.28423  &0.323103 \\
$^{24}$Na & -1.59925 &0.3403    &1.92878  &0.289744 &
$^{25}$Na & -5.01558 &0.302793  &5.32092  &0.287093 \\
$^{26}$Na & -4.65003 &0.297148  &4.94264  &0.280147 &
$^{27}$Na & -5.93258 &0.398734  &6.37662  &0.369386 \\
$^{28}$Na & -2.01555 &0.388844  &2.46011  &0.324078 &
$^{29}$Na & -1.57596 &0.361364  &2.01826  &0.297191 \\
$^{30}$Na & -3.8188  &0.315588  &4.2291   &0.291141 &
$^{31}$Na & -5.99753 &0.27771   &6.3129   &0.263971 \\
$^{32}$Na & -2.84109 &0.216843  &3.11021  &0.203505 &
$^{33}$Na & -2.05982 &0.217242  &2.3596   &0.200594 \\
$^{34}$Na & -0.815459&0.235539  &1.12203  &0.193029 &
$^{35}$Na & -0.440892&0.262169  &0.748431 &0.182956 \\
$^{36}$Na & -0.425172&0.257764  &0.729539 &0.178082 &
$^{37}$Na & -0.426312&0.252979  &0.725345 &0.174057 \\
$^{38}$Na & -0.560764&0.236309  &0.85135  &0.174332 &
$^{39}$Na & -0.767664&0.221727  &1.04873  &0.174823 \\
\hline
$^{22}$Mg & -3.5102   &0.337634  &3.86364  &0.315119 &
$^{23}$Mg & -4.68847  &0.341192  &5.03992  &0.32147  \\
$^{24}$Mg & -5.58442  &0.345115  &5.93632  &0.326273 &
$^{25}$Mg & -5.70374  &0.32602   &6.03196  &0.308936 \\
$^{26}$Mg & -5.91305  &0.308559  &6.21652  &0.293159 &
$^{27}$Mg & -6.04836  &0.305345  &6.33983  &0.289534 \\
$^{28}$Mg & -6.28349  &0.303037  &6.56421  &0.287016 &
$^{29}$Mg & -2.13344  &0.407667  &2.57616  &0.332949 \\
$^{30}$Mg & -3.19583  &0.260324  &3.48083  &0.240888 &
$^{31}$Mg & -4.19839  &0.228173  &4.44081  &0.216497 \\
$^{32}$Mg & -2.65606  &0.288664  &2.97542  &0.256755 &
$^{33}$Mg & -2.40436  &0.311802  &2.77146  &0.270189 \\
$^{34}$Mg & -1.27359  &0.233869  &1.56893  &0.201356 &
$^{35}$Mg & -2.06917  &0.231846  &2.37601  &0.209307 \\
$^{36}$Mg & -0.808364 &0.255092  &1.1172   &0.197461 &
$^{37}$Mg & -0.700842 &0.25534   &1.00767  &0.191015 \\
$^{38}$Mg & -0.779685 &0.247165  &1.07905  &0.188717 &
$^{39}$Mg & -0.944354 &0.236374  &1.23856  &0.187856 \\
$^{40}$Mg & -1.6618   &0.220155  &1.94604  &0.191281 &&&&&\\ 
\hline 
$^{24}$Al & -1.45476 &0.340458  &1.78352  &0.286481 &
$^{25}$Al & -5.33616 &0.325916  &5.6647   &0.30783  \\
$^{26}$Al & -8.96115 &0.414522  &9.43319  &0.393303 &
$^{27}$Al & -4.25196 &0.384866  &4.61596  &0.346254 \\
$^{28}$Al & -9.60618 &0.387342  &10.0265  &0.368222 &
$^{29}$Al & -10.1518 &0.387245  &10.5648  &0.367966 \\
$^{30}$Al & -8.57763 &0.358694  &8.99408  &0.340275 &
$^{31}$Al & -3.93913 &0.363267  &4.39077  &0.326436 \\
$^{32}$Al & -2.62455 &0.320816  &3.00351  &0.280775 &
$^{33}$Al & -1.90801 &0.223972  &2.13351  &0.199603 \\
$^{34}$Al & -1.76454 &0.22558   &2.02282  &0.200253 &
$^{35}$Al & -4.40195 &0.215828  &4.66863  &0.205308 \\
$^{36}$Al & -4.45984 &0.219293  &4.74574  &0.208377 &
$^{37}$Al & -4.54495 &0.220995  &4.83685  &0.209645 \\
$^{38}$Al & -4.7981  &0.217558  &5.07953  &0.206567 &
$^{39}$Al & -4.69265 &0.214125  &4.97399  &0.203054 \\
$^{40}$Al & -3.60492 &0.225556  &3.88433  &0.208409 &
$^{41}$Al & -3.50796 &0.222836  &3.78546  &0.205275 \\
$^{42}$Al & -2.99132 &0.248402  &3.32657  &0.222474 &
$^{43}$Al & -3.05889 &0.253024  &3.39251  &0.22516  \\
$^{44}$Al & -3.02871 &0.249435  &3.36787  &0.221941 &
$^{45}$Al & -2.9738  &0.260693  &3.31665  &0.228223 \\
$^{46}$Al & -3.1057  &0.245182  &3.43189  &0.217474 &
$^{47}$Al & -0.319467&0.289022  &0.63011  &0.153346 \\
\hline 
$^{26}$Si & -7.91999 &0.414491  &8.39401  &0.390892 &
$^{27}$Si & -4.17883 &0.381022  &4.53972  &0.342842 \\
$^{28}$Si & -3.26289 &0.341632  &3.63435  &0.30699  &
$^{29}$Si & -3.778   &0.363259  &4.1903   &0.326487 \\
$^{30}$Si & -6.2779  &0.265629  &6.52777  &0.253178 &
$^{31}$Si & -2.90607 &0.336128  &3.28085  &0.294991 \\
$^{32}$Si & -2.94072 &0.303651  &3.28119  &0.270526 &
$^{33}$Si & -1.99993 &0.20584   &2.19687  &0.186214 \\
$^{34}$Si & -2.06805 &0.212795  &2.27827  &0.191571 &
$^{35}$Si & -2.00258 &0.217443  &2.24468  &0.195173 \\
$^{36}$Si & -2.08563 &0.206827  &2.29686  &0.185987 &
$^{37}$Si & -2.02117 &0.221807  &2.28665  &0.197815 \\
$^{38}$Si & -2.03642 &0.226282  &2.31387  &0.200671 &
$^{39}$Si & -2.14661 &0.225337  &2.41648  &0.199744 \\
$^{40}$Si & -3.18035 &0.30413   &3.59945  &0.266638 &
$^{41}$Si & -2.98979 &0.250094  &3.33998  &0.225143 \\
$^{42}$Si & -2.72873 &0.236071  &3.07376  &0.21272  &
$^{43}$Si & -2.60821 &0.230403  &2.95853  &0.20749  \\
$^{44}$Si & -2.87398 &0.234379  &3.22239  &0.211086 &
$^{45}$Si & -6.43192 &0.225856  &6.78428  &0.215118 \\
$^{46}$Si & -3.39586 &0.244309  &3.75371  &0.219991 &
$^{47}$Si & -3.19051 &0.231893  &3.5316   &0.208824 \\
$^{48}$Si & -0.357803&0.322425  &0.671824 &0.159255 & &&&& \\
\hline \hline
\end{tabular}
\label{tab1}
\end{table*}
%%%%%%%%%%%%%%%%%%%%%%%%%%%%%%
\begin{figure}[H]
\centering
\includegraphics[width=0.5\textwidth]{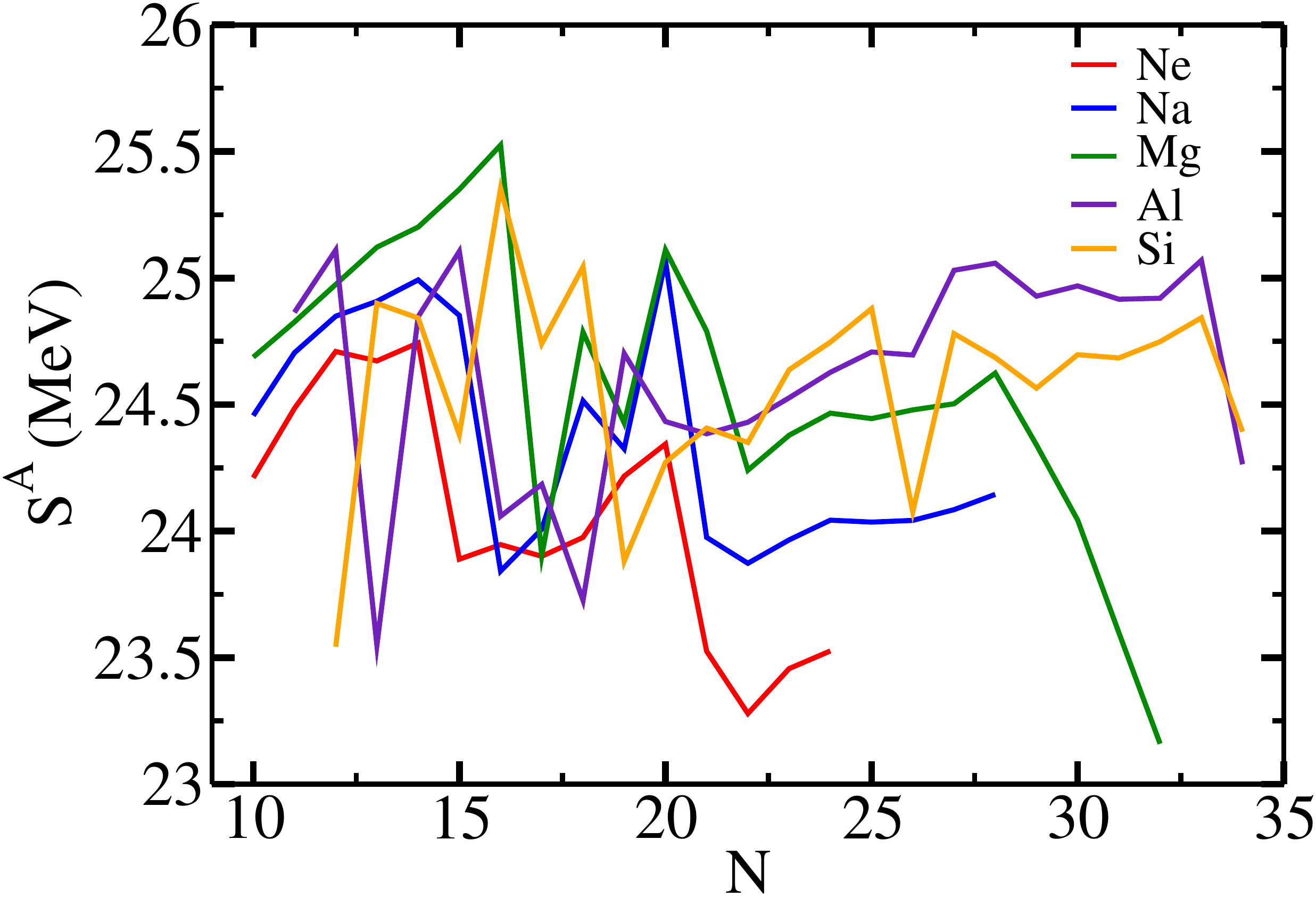}
%\vspace{0.1 cm}
\caption{(color online) The symmetric energy is shown as a function of neutron number N for Ne, Na, Mg, Al and Si nuclei.}
\label{S^{A}}
\end{figure}
\begin{figure}[H]
\centering
\includegraphics[width=0.5\textwidth]{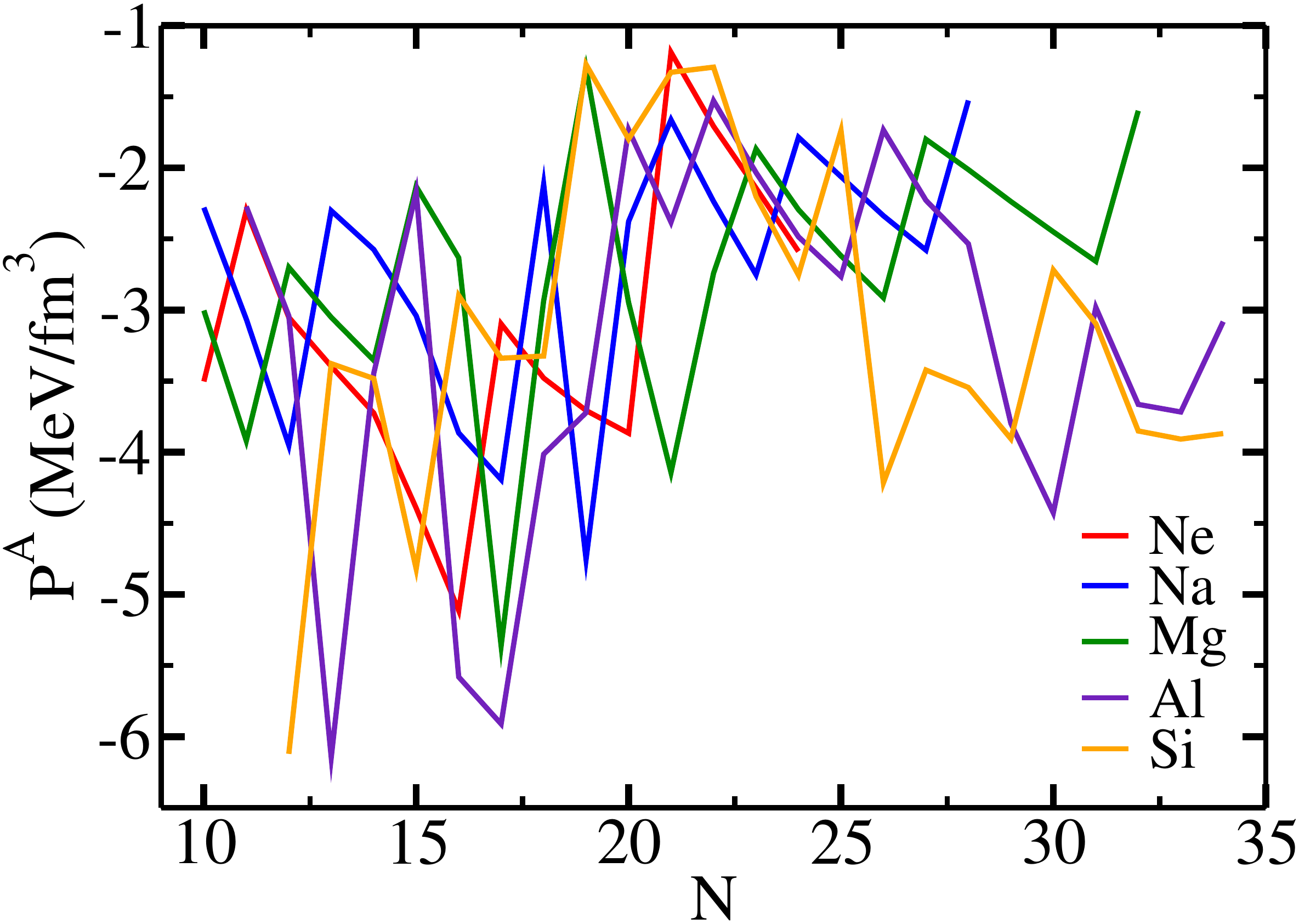}
%\vspace{0.1 cm}
\caption{(color online) Here we have represented neutron pressure as a function of neutron number N for Ne, Na, Mg, Al and Si nuclei.}
\label{P^{A}}
\end{figure}
%%%%%%%%%%%%%%%%%%%%
We have shown the equivalent density as a function of radial coordinate for $^{20,29}$Ne, $^{23,31}$Na, $^{24,39}$Mg and  $^{28,45}$Si isotopes as representative cases. It is to be noted that the density plots in Fig. \ref{diffuse} reflect the normalized equivalent density of the deformed nucleus. In each panel, we have shown the equivalent densities for the isotopes from the $\beta-$stable and neutron drip-line region for each atomic nucleus, which provides the information relative changes of the density with-respect-to isospin asymmetry. For example, panel (a), (b), (c), and (d) are assigned for the isotopes of $^{20,29}$Ne, $^{23,31}$Na, $^{24,30}$Mg, and $^{28,45}$Si, respectively. From the spherical equivalent density, we obtained the weight function $\vert F(x) \vert^2 $ of the nucleus. This is one of the most important ingredients to determine the surface properties of the nucleus. In each specific representation, although the small difference in the surface region of the densities reflects a wide difference in the weight functions. As a consequence, one can expect a very different surface property for $\beta-$stable and drip-line nuclei.
\begin{figure}[H]
\centering
\includegraphics[width=0.5\textwidth]{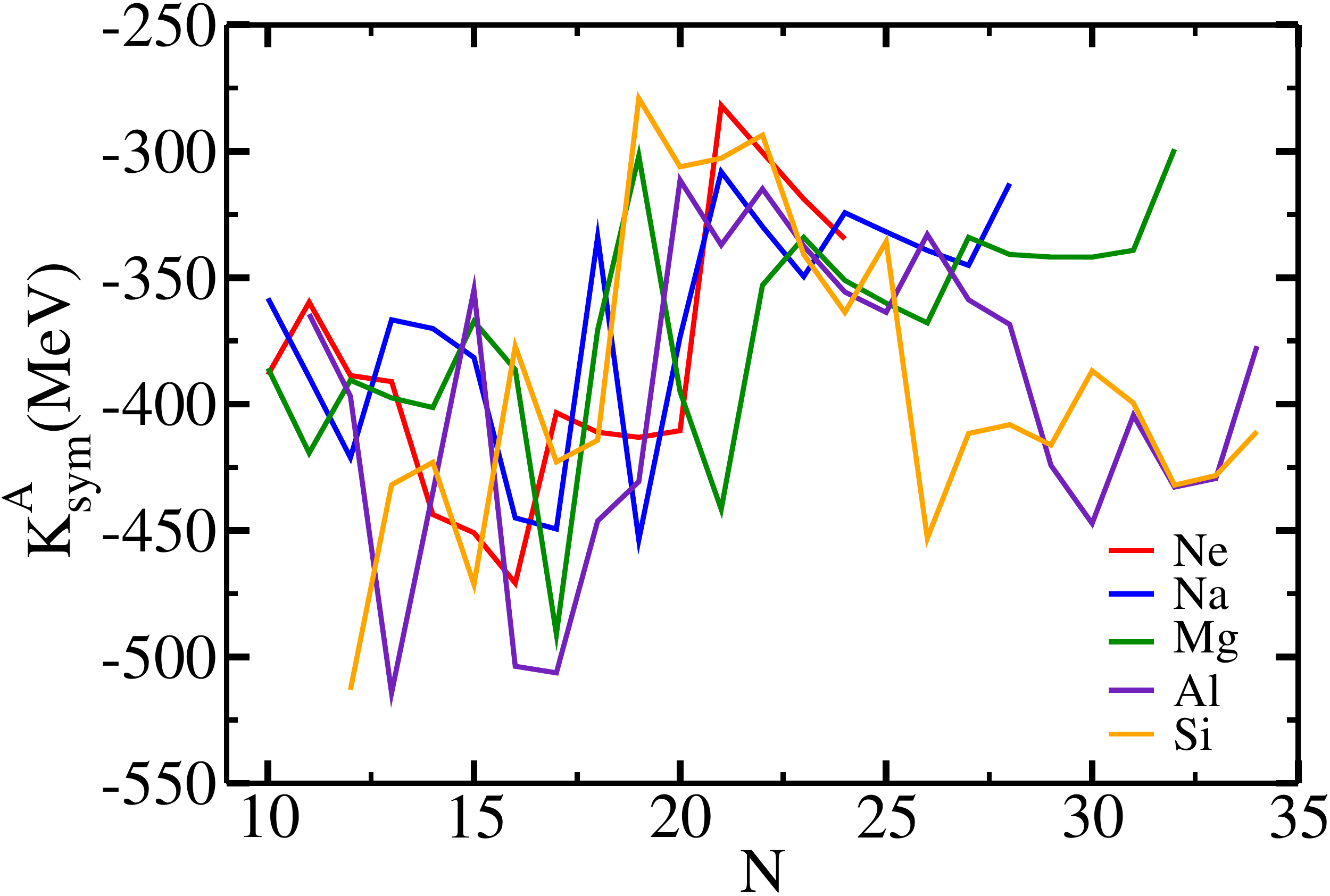}
%\vspace{0.1 cm}
\caption{(color online) The symmetry energy curvature coefficients are given for the Ne, Na, Mg, Al and Si nuclei isotopic series.}
\label{K_{sym}^{A}}
\end{figure}
%%%%%%%%%%%%%%%%%
\begin{figure}[H]
\centering
\includegraphics[width=1.05\columnwidth]{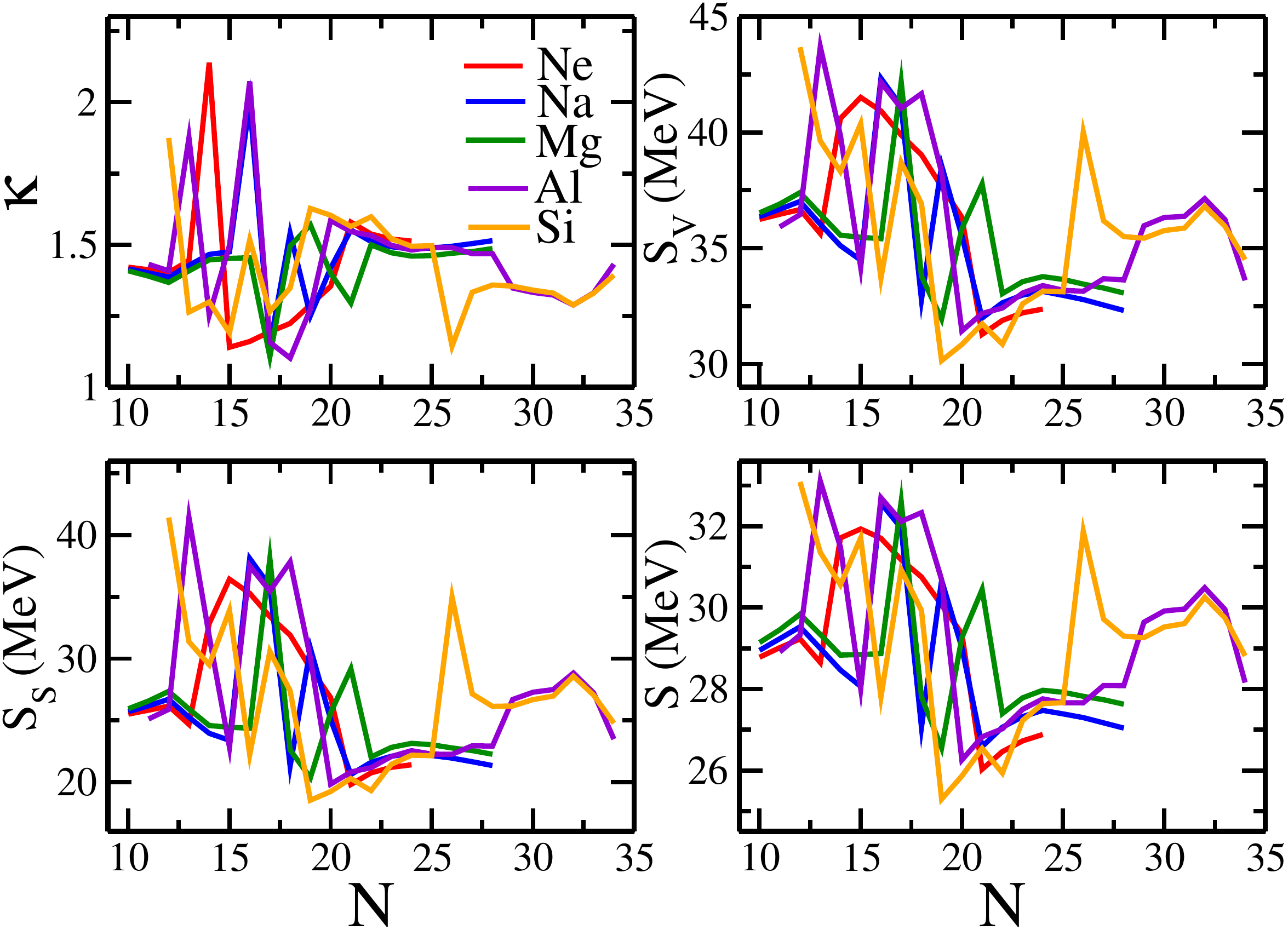}
%\vspace{0.1 cm}
\caption{(color online) The volume ($S_V$), surface ($S_s$) and total surface energy (S) are shown. The $\kappa-$factor is also given as a function of neutron number N for Ne, Na, Mg, Al and Si nuclei.}
\label{eosden}
\end{figure}
%%%%%%%%%%%%%%%%%%%%%%%%%%%%%%%%%%%%%%%%%%

\subsection{Nuclear surface properties}
There are several prescriptions to analyze nuclear surface properties of finite nuclei in terms of isospin dependent nuclear matter quantities such as symmetry energy and its derivatives. In the present work, we have mainly highlighted the CDFM approach of Br\"uckner's functional and the liquid drop approximation of Danielewicz \cite{dani03}. Furthermore, the symmetry energy is separated into it's volume and surface components, which are mainly approximated in connection with the nuclear matter's properties at the saturation. For example, Eq. (\ref{s0t}) shows that the saturation symmetric energy $S(\rho_0)$ and nuclear matter density are needed to calculate the symmetric energy of a finite nucleus. On the other hand, for the same quantity, we need in the nuclear matter Br\"uckner's functional as of Eqs. (\ref{s0})$-$(\ref{k0}). The calculated nuclear surface properties such as symmetric energy, neutron pressures, and symmetry energy curvature are given in Fig. \ref{S^{A}}, \ref{P^{A}}, and \ref{K_{sym}^{A}}, respectively as a function of the neutron number. In each figure, we have given the corresponding calculated quantities for the isotopic chain of considered nuclei Ne, Na, Mg, Al, and Si. From the Fig. \ref{S^{A}}, it can be noticed that the symmetry energy for all the isotopes bounded within a range of 23-25 $MeV$.  It is worth mentioning that the exotic nuclei having large neutron-to-proton asymmetry respond to the nature of nuclear symmetry energy. Further, the appropriate information of symmetry energy from finite nuclei will be added to a wide range of nuclear phenomena starting from the study of nuclear structure, dynamics of heavy-ion reactions to the high isospin asymmetry system such as neutron star matter. 

Observing the characteristics of symmetry energy $S^{A}$, and surface symmetry curvature $K_{sym}^{A}$ in Figs. \ref{S^{A}}, and \ref{K_{sym}^{A}}, in general, distorted peaks appear at N = 14, 16, 20, and 28, for some of the cases. Similarly, we also notice a fall in the neutron pressure $P^{A}$ at the same neutron numbers in Fig. \ref{P^{A}}. We found the value of $S^{A}$ within the range of 24-25 $MeV$ at the neutron number N = 20 for all the isotopic series of Ne, Na, Mg, Al, and Si nuclei. In all the figures, the trend is not smooth, this is either from the transform of deformed density into the spherical equivalent one by using the two Gaussian fittings (discussed in sub-section \ref{equiv}) or by neglecting the shape degrees of freedom. Although there is an anomalous trend appear in all the isotopic chain, a careful inspections show that the magnitude of the peak and/or depth are assuredly at the neutron magic N =20 and also for 28 along with a few more neutron numbers, namely N = 14, and 16. These larger (lower) magnitude of $S^{A}$, $P^{A}$ and $K_{sym}^{A}$ provides the signature of shell/sub-shell closures over the isotopic chain. It is to be noted that, one can obtain the $K_{sym}^{A}$ by adopting the Thomas-Fermi approach along with relativistic mean-field theory within the leptodermus expansion for finite nuclei \cite{patra2002curv}. Here the neutron pressure and symmetry energy curvature are distributed within a range of -7 to -1 $MeV/fm^{3}$ and -550 to -250 $MeV$ respectively. In some of our earlier study \cite{bhu18,abdul19,manpreet}, we notice a relation of $S^{A}$ with the shell closure nature of magic number, i.e., we get a maximum of $S^{A}$ at the magic number. This correlation is seen only for light and medium mass nuclei. While searching such relation in the considered mass region although, we noticed such type of peaks at the presently reported magic number \cite{plb2020Han} within the Br\"uckner's functional, but it does not show a regular pattern, as we are getting in medium mass region \cite{manpreet}.
%%%%%%%%%%%%%%%%%%%%%%%
For further analysis of the surface properties of symmetry energy, we subdivided the $S$ values into its surface and volume components as shown in Fig. \ref{eosden}. To calculate these quantities, we have used the prescription of P. Danielewicz \cite{dani03,dani04,dani06,dani09}. The estimation of $S$ with Eq. (\ref{s0t}) is quite different than that of (\ref{s0}) \cite{dani03,dani04,dani06,dani09}. As a result, we predict nearly two different values of $S$ in both two approaches, which can be seen clearly in Figs. (\ref{S^{A}})  and (\ref{eosden}). In this Fig. \ref{eosden} we have produced the volume, the surface, and the total surface energy. The parameter $\kappa = S_V/S_S$, i.e., the ratio of the volume to the surface part of the symmetric energy is also given in the figure. We observed that the symmetry energy related to volume is more than the surface contribution and their ratio $\kappa$ decreases with increasing of N. As the symmetric energy is pretty much related to N-Z asymmetry, hence it is crucial to examine the surface and volume part. At the surface of the nucleus, the nuclear saturation does not acquire, and also due to the unequal population of the number density of the nucleons, the surface symmetry energy get more crucial, detail explanation can be found in Ref. \cite{feenberg1947}. 
%%%%%%%%%%%%%%%%%%%%%

\section{Concluding Remarks}
\label{summary}
In our outline, the properties of the surface of a nucleus, such as symmetric energy, neutron pressure, and the symmetry energy curvature coefficient as a function of the neutron number are investigated for Ne, Na, Mg, Al, and Si nuclei from the proton to neutron drip-lines. The Coherent Density Fluctuation Model (CDFM) is used to evaluate these quantities within the framework of the relativistic mean-field formalism. Here, in CDFM the nuclear/nucleonic density does not follow any sharp edge surface; thus, the diffuseness parameter is also not neglected. The fluctuation of the flucton attains infinitesimal size/dimension and is distributed over the range of nuclear density distribution from the relativistic mean-field model. Hence, there is no possibility of sloppy surface contribution from the density in terms of weight function. And further, we have examined the CDFM account the surface effect substantially compared to Liquid-Drop-Approximation \cite{subrat} and reference therein. We have discussed the bulk properties of these isotopic chain, which are essential for determining the symmetry energy and its co-efficient. We evaluated the binding energy, the binding energy per particle, charge radius, nuclear quadruple moment deformation parameter, and the neutron-skin thickness. To locate the recently reported magic structure of some of the isotopes in the considered region, we analyzed the two-neutron separation energy. The correlation of the magic number with the surface symmetric properties of the isotopes are studied. We noticed some relation in the reported isotopes, which is recently a fact. As expected, the neutron-skin thickness increases monotonously with neutron numbers in an isotopic series. On the other hand, the BE/A increases initially and then decreases with N. The shape of the nucleus in an isotopic chain changes and does not show any regular pattern. This is because a larger number of nuclei exhibit shape coexistence. 

We converted the deformed densities to their spherical equivalent with a two Gaussian fitting. The coefficients $a_1$ and $a_2$ are in general decrease, but $c_1$ increases in an zig-zag manner. The values of $c_2$ however decrease with neutron number. In practice, when we use the Gaussian fitting for the axial density distribution to obtain the spherical one, there is an enhancement of central density with a long tail. As a result, the surface covers a wide range of the radius, which is the limitation of the Gaussian fitting procedure. In some of the previous studies by our collaborators, various fitting techniques were used to convert axial deform density to spherical form and their effect on the surface properties of finite nuclei quantitatively \cite{kishore20} and reference therein.

In our present work, we reported here two types of symmetry energies, one due to liquid drop approximation and the other due to LDA by Br\"uckner method. Although we can not measure the symmetric energy often directly, we can extract the values from other nuclear observable, indirectly which are related to it \cite{bhu18}. But the question is why we are not getting the same values or the same trends of symmetry energy for two different approximations? Although the dimensions are the same in both approaches the expressions are different and the fact is that these nuclei are small in size. Nuclei having large size show almost the same values of symmetry energy in both approximations as we have shown in our previous work \cite{quddus2020}. Here we get the maximum value of symmetry energy in the liquid drop approach which is $33.11$ and $25.52$ $MeV$ in Br\"uckner functional adoption for NL3* parameter set.

In the earlier studies, it is shown that the symmetric energy and/or neutron pressure is connected with the shell/sub-shell closure in the emergence of a peak and/or fall over an isotopic chain \cite{gad11,gad12,bhu18,quddus2020}. Hence the symmetry energy and its co-efficient (neutron pressure) can be used as an observable for predicting the neutron and/or proton magic at the drip-line region. Although we get considerable magnitude in the symmetric energy for these neutron numbers (N=14, 16 and 32/34), but a broke-down/ distorted trend in the isotopic chain of all the considered nuclei. The abnormal trend in these mass regions can be correlated with the possible reasons:
\begin{itemize}
\item The considered nuclei in the present study are lighter mass nuclei, hence quite sensitive to the neutron-to-proton ratio. Analogous to the neutron proton ratio, the neutron-to-proton difference is very small, hence the isospin dependency of the system not effective, as a result, the symmetry energy for these mass nuclei is not followed the systematic trend.
\item The considered mass nuclei are found to be highly deformed in their ground state, which reflects in their densities. Following the limitation of the calculation, the spherical equivalent densities using a slender approximation are implemented into the CDFM to obtain the weight function. It is worth mentioning that, the density in terms of weight function is one of the central ingredients for obtaining the symmetry energy \cite{kishore20} and its components.
\item As we mentioned above, the considered isotopes are small in size, hence it is likely to get clustering configuration with sharp surface instead of a smooth surface as that of nuclei of mass A $\geq$ 40 \cite{arima72,aremugumpatra05,bhu13,nora08}, it has been shown that the smaller nuclei in the mass table show cluster properties unlike developing a smooth surface as that of medium or heavy mass nuclei. 
\end{itemize}
Therefore we suggest, while doing these further calculations in the lower mass region, one must need to take care of these points.

\noindent 
{\bf  Acknowledgement:} One of the authors (JAP) is thankful to the Institute of Physics, Bhubaneswar for providing the computer facilities during the work. This work is partly reinforced by SERB, Department of Science and Technology, Govt. of India, Project No. CRG/2019/002691. MB acknowledges the support from FAPESP Project No. 2017/05660-0, FOSTECT Project No. FOSTECT.2019B.04, and the CNPq - Brasil. \\
%%%%%%%%%%%%

%\bibliography{cdfm.bib}
%\bibliographystyle{apsrev4-1}
%\begin{comment}

%\end{comment}

\end{document}